\documentstyle[12pt]{article}
\begin{document}

\title{Connection between closeness of classical orbits and the 
factorization of radial Schr\"{o}dinger equation}

\author{Y. F. Liu, W. J. Huo and J. Y. Zeng\\Department of Physics, Peking University
, Beijing, China}
 
\baselineskip 25pt

\maketitle

\begin{abstract}
  It was shown that the Runge-Lenz vector for a hydrogen atom is equivalent
   to the raising and lowering operators derived from the factorization of 
   radial Schr\"{o}dinger equation.  Similar situation exists for an isotropic 
   harmonic oscillator.  It seems that there may exist intimate relation 
   between the closeness of classical orbits and the factorization of radial
   Schr\"{o}dinger equation.  Some discussion was made about the factorization
    of a 1D Schr\"{o}dinger equation. 

\vspace{1.0in}{\bf PACS} number(s), 03.65.-w, 03.65.Fd, 03.65.Ge
\end{abstract}
{\center \section*{I. INTRODUCTION}}
~~
\par
     The factorization method of Schr\"{o}dinger $^{[1,2]}$ was
     extended to address the radial Schr\"{o}dinger equation of a
     particle in a central potential $V(r)^{[3-6]}$. It was shown$^{[6]}$
     that only for two kinds of potential, the Coulomb potential and
     isotropic harmonic oscillator, the radial Schr\"{o}dinger equation
     can be factorized, and directly from the factorization of radial
     equation,  {\sl two kinds} of raising and lowering operators, 
     $A_{\pm}(l)$ and $B_{\pm}(l)$, can be derived for an isotropic harmonic
     oscillator, but {\sl only one}, $A_{\pm}(l)$, for a hydrogen atom. 
     This reminds us of the famous Bertrand theorem$^{[\rm 7]}$ in classical
     mechanics, which says: ``the only central forces that result in close
     orbits for all bound particles are the inverse square law and Hooke's
     law".  It is guessed that the factorizability of radial Schr\"{o}dinger
     equation may be intimately connected with the closeness
     of classical orbits. 

     In classical mechanics, the maximum number of functional independent 
     conserved quantities of a closed Hamiltonian system with $N$ degrees
     of freedom is $2N-1^{[8]}$.  For a system with independent conserved
     quantities no less than $N$ is called integrable$^{[\rm 9]}$. 
     An integrable classical system with $N+\Lambda$ independent conserved
     quantities $(0\le \lambda\le N-1)$ is called
     $\Lambda$-fold degenerate, and there exist $\Lambda$ linear relations
     with integer coefficients between the $N$ frequencies $\omega_i
     (i=1, 2, \cdots, N)$ of the system$^{[10]}$.  A classical system for
     $\Lambda=N-1$ is called a completely degenerate system, and there remains
     only one independent frequency.  For example, it is well-known that for a particle
     in a central potential $V(r)$, apart from the Hamiltonian, the angular
     momentum $\bf l$ is also conserved, and the particle in a general central
     potential $V(r)$ is $1$-fold degenerate and moves in a plane, 
     but the planar orbits are in general not closed.  For a particle
     in an attractive Coulomb potential $(V(r)=-\kappa/r)$, it was
     demonstrated$^{[11]}$ that there exists an additional conserved quantity, 
     the Runge-Lenz vector, $(m=\kappa=1)$, ${\bf{a=p \times l-r}/\mit{r}}$, 
     which guarantees the closeness of a Kepler orbit.  In fact,  the direction
     of $\bf{a}$ \rm is just that of the major axis of the elliptic orbit and
     the magnitude of $\bf{a}$  is the eccentricity.  The existence of
     Runge-Lenz vector implies that the Coulomb potential has a higher
     dynamical symmetry $SO_{4}$ than its geometric symmetry $SO_{3}$ $^{[12]}$. 
     However,  it is seen that ${\bf a\cdot l}=0$, and ${\bf a}^2=2{\mit H}{\bf l}^2+1$, 
     so the number of independent conserved quantities is 5,  and the hydrogen
     atom is a completely degenerate system.  Similar situation exists for an
     isotropic harmonic oscillator. 

     In Sect. II it will be shown that for a 2-dimensional(2D) hydrogen atom,  
     the Runge-Lenz
     vector is equivalent to the raising and lowering operators derived from
     the factorization of radial Schr\"{o}dinger equation. In Sect. III, 
     it will be shown that,  for a 3D hydrogen atom, from the Runge-Lenz vector
     $\bf{a}$ and angular momentum $\bf{l}$, one can construct three kinds
     of raising and lowering operators $(\Delta n=0, \Delta l=\pm1, \Delta
     m=0, \pm1)$, which are equivalent to the raising and lowering operators
     derived from the factorization of radial Schr\"{o}dinger equation. 
      In Sect. IV we will address isotropic harmonic oscillators. 
      In terms of the two kinds of raising and
     lowering operators one can construct the conserved quantities
     characterizing elliptic orbits. Finally, in Sect. V there is some discussion
     about the factorizability of the Schr\"{o}dinger equation for a 1D system,  
     which has been investigated extensively in supersymmetric quantum mechanics. 
{\center \section*{II. 2D HYDROGEN ATOM}}  
   \subsection*{(A) Runge-Lenz vector and a simple algebraic approach to
   the eigenvalue problem}
~~
\par
     For a 2D hydrogen atom, the quantum version of the Runge-Lenz vector
     reads  $(\hbar=m=e=1)$ 
        \begin{equation} 
         {\bf a}=\frac{1}{2}({\bf p}\times {\bf l-l} \times{\bf p})-
         {\bf e}_{\rho}={\bf p \times l-{\mit i}p}-{\bf e}_{\rho} ,
        \end{equation}
     where ${\bf p}={\mit p_{x}}{\bf i}+{\mit p_{y}}{\bf j} , {\bf l}
     ={\mit l_{z}}{\bf k}={\mit(xp_{y}-yp_{x}){\bf k} ,  {\mit \rho=\sqrt{x^{2}+y^{2}}}}$. 
     It is easily shown that
        \begin{equation} 
         [l_{z}, a_{x}]=ia_{y}, [l_{z}, a_{y}]=-ia_{x},  
         [a_{x}, a_{y}]=-i2Hl_{z} . 
        \end{equation}
     In the subspace spanned by the bound states with eigenenergy $E(E<0), a_{x}$ 
     and $a_{y}$ may be replaced by $A_{x}=\sqrt{-1/2E} a_{x}, A_{y}=\sqrt
     {-1/2E} a_{y}$.  Let $A_{z}=l_{z}$, then
        \begin{equation} 
         [A_{\alpha}, A_{\beta}]=i\varepsilon_{\alpha\beta\gamma}A_{\gamma}  ,
        \end{equation}
     i.e. $A_{x}, A_{y}$ and $A_{z}$ constitute the $SO_{3}$ Lie algebra, and
     the eigenvalue of ${\bf A}^{2}=-\frac{1}{4}-\frac{1}{2{\mit E}}$
     is $m_{0}(m_{0}+1), m_{0}=0, 1, 2, \cdots$. Therefore, the energy eigenvalue
     of a 2D hydrogen atom is 
        \begin{equation} 
          E=E_{n}=-1/2n^{2}, n=(m_{0}+1/2)=1/2, 3/2, 5/2, \cdots .
        \end{equation}

     Alternatively, defining $a_{\pm}=a_{x}\pm ia_{y}$, it is easily shown that
        \begin{equation} 
          [l_{z}, a_{\pm}]=\pm a_{\pm}   ,
        \end{equation}
        \begin{equation} 
          a_{-}a_{+}=\frac{H}{2} (2l_{z}+1)^{2}+1 ,
        \end{equation}
     i.e. $a_{\pm}$ are just  the  raising and  lowering operators  of the 
     magnetic  quantum number $m$. Assume $|Em\rangle$ is the eigenstate of 
     $(H, l_{z})$ with eigenvalues $(E, m)$, then  $a_{\pm}|Em\rangle$ are  also
     the  eigenstates of $H$ with energy $E$ and the   eigenstates  of $l_{z}$
     with eigenvalues $(m \pm 1)$. For a given energy eigenvalue $E_{n}$, the
     allowed $|m|$ must have an upper limit $m_{0}$, and $a_{+}|Em_{0}\rangle=0$.
     Hence$\quad a_{-}a_{+}|Em_{0}\rangle=0$. Using (6), we get $\frac{E}{2} (2m_{0}+1)^
     {2}=1$, which  is just (4). The degenerate states belonging to $E_{n}$ may
     be expressed as  $a^{k}_{-}|E_{n}m_{0}\rangle, k=0, 1, \cdots, 2m_{0}$, and  the
     degeneracy is $2n=(2m_{0}+1)=1, 3, 5, \cdots.$

  \subsection*{(B) Raising  and lowering operators derived from factorization}
  ~~
\par
     The energy eigenstate of a 2D hydrogen  atom  may be chosen  as the
     simultaneous eigenstate of $(H, l_{z})$, i.e. $\Psi(\rho, \phi)={\rm e}^
     {im\phi}\chi _{m}(\rho)/\sqrt{\rho}, m=0, \pm 1, \pm 2, \cdots$, and
     $\chi_{m}(\rho)$ satisfies $H(m)\chi_{m}(\rho)=E\chi_{m}(\rho)$, or
       \begin{equation} 
         D(m)\chi_{m}(\rho)=\lambda_{m}\chi_{m}(\rho), \lambda_{m}=-2E , 
       \end{equation}
       \begin{displaymath}
         D(m)=-2H(m)=\frac{d^2}{d\rho^2} - \frac{m^{2}-1/4}{\rho^{2}}
          + \frac{2}{\rho}  .
       \end{displaymath}
     Directly   from the factorization one may derive the raising and
     lowering operators$^{[6]}$,
       \begin{displaymath}
        A_{+}(m)=\frac{d}{d\rho}-\frac{m+1/2}{\rho}+\frac{1}{m+1/2} ,
       \end{displaymath}
       \begin{equation}
        A_{-}(m)=\frac{d}{d\rho}+\frac{m-1/2}{\rho}-\frac{1}{m-1/2}  ,
       \end{equation}
    whose selection rules are $\Delta E=0(\Delta n=0) $ and $\Delta m=\pm 1$.
    Using $A_{\pm}(m)$, the energy eigenvalues and eigenstates can also be
    easily obtained. From (7) it  is seen that  $E$ depends  only on the 
    absolute value of $m$. Using  Hellmann-Feynman theorem
       \begin{equation}
         \frac{\partial E}{\partial |m|}=\langle\frac{\partial
          H(m)}{\partial |m|}\rangle=\frac{|m|}{\rho^{2}}>0 ,
       \end{equation}
    i.e. $E$ increases monotonically with $|m|$. Thus, $m=0$ for the ground 
    state. On the other hand, for a given $E, |m|$ must have an upper limit,  
    say, $m_{0}( m_{0}>0)$ . Then
       \begin{equation}
         A_{+}(m_{0})\chi_{m_{0}}(\rho)=(\frac{d}{d\rho}-\frac{m_{0}+1/2}
         {\rho}+\frac{1}{m_{0}+1/2})\chi_{m_{0}}(\rho)=0 .
       \end{equation}
    So  $\chi_{m_{0}}(\rho) \sim \rho^{m_{0}+1/2}{\bf e}^{-\rho/(m_{0}+1/2)}$.
    From (10) we have $A_{-}(m_{0}+1)A_{+}(m_{0})\chi_{m_{0}}(\rho)=0$,
    and using (7) and (8), we get
       \begin{equation}
        [D(m_{0})-1/(m_{0}+1/2)^{2}]\chi_{m_{0}}(\rho)=[-2E_{m_{0}}-1/(m_{0}
        +1/2)^{2}]\chi_{m_{0}}(\rho)=0 .
       \end{equation}
    Thus we get $E_{m_{0}}=-1/2(m_{0}+1/2)^{2}$. Using $A_{-}(m_{0}),
    A_{-}(m_{0}-1), \cdots$, successively operating on $\chi_{m_{0}}(\rho)$,
    one may get all the degenerate eigenstates belonging to $E_{m_{0}},
    \chi_{m_{0}, m}(\rho), m=m_{0}, m_{0}-1, \cdots, -m_{0}$. Alternatively, the
    degenerate eigenstates belonging to $E_{n}=-1/2n^{2}, (n=1/2, 3/2, 5/2, 
    \cdots)$ may be denoted by $\chi_{n, m}(\rho), |m|=n-1/2, n-3/2, \cdots, 1, 0$, 
    and it can be shown that $\chi_{n, m}(\rho) \sim \rho^{|m|+1/2}{\bf e}^
    {-\rho/n}F(-n_{\rho}, 2|m|+1, \frac{2\rho}{n})$, where $F$ is the confluent
    hypergeometric function, and $n_{\rho}=(n-1/2)-|m|=0, 1, 2, \cdots, (n-1/2)$.

 \subsection*{(C) Equivalence of the Runge-Lenz vector and
  the raising and lowering operators}
~~
\par
    In polar coordinates the Runge-Lenz vector (1) can be expressed as
    $a_{\pm}=a_{x} \pm ia_{y}$
       \begin{equation}
         a_{\pm}={\rm e}^{\pm i\phi}[(\mp \frac{\partial}{\partial \rho}
         +\frac{1}{\rho}l_{z})l_{z}-\frac{1}{2}(\frac{\partial}{\partial \rho}
         \mp \frac{1}{\rho}l_{z})-1] .
       \end{equation}
    Operating on the eigenfunction ${\rm e}^{im\phi}R_{m}(\rho)$, the raising
    and lowering of $m$ in the angular function are accomplished by ${\rm e}^
    {\pm i\phi}$, and $a_{\pm}$ are equivalent to operators $a_{\pm}(m)$
    operating on $R_{m}(\rho)$, 
       \begin{eqnarray} 
         a_{\pm}(m)&=&[(\mp m\frac{\partial}{\partial \rho}+\frac{m^{2}}{\rho})
         -\frac{1}{2}\frac{\partial}{\partial \rho} \pm \frac{m}{2\rho}-1]
         \nonumber \\
                   &=&-(\frac{1}{2} \pm m)\frac{\partial}{\partial \rho}+
                \frac{m(m \pm 1/2)}{\rho}-1 .
        \end{eqnarray}
    Apart from a trivial constant factor, $a_{\pm}(m)$ may be expressed as
       \begin{displaymath}
         a_{+}(m)=\frac{\partial}{\partial \rho}-\frac{m}{\rho}+
         \frac{1}{m+1/2} , 
       \end{displaymath}
       \begin{equation}
         a_{-}(m)=\frac{\partial}{\partial \rho}+\frac{m}{\rho}-
         \frac{1}{m-1/2} , 
       \end{equation}
    The operators $a_{\pm}(m)$ operating on $R_{m}(\rho)$ may be replaced
    by $A_{\pm}(m)$ operating on $\chi_{m}(\rho)=\sqrt{\rho}R_{m}(\rho)$
       \begin{displaymath}
         A_{+}(m)=\frac{\partial}{\partial \rho}-\frac{m+1/2}{\rho}+
         \frac{1}{m+1/2} , 
       \end{displaymath}
       \begin{equation}
         A_{-}(m)=\frac{\partial}{\partial \rho}+\frac{m-1/2}{\rho}-
         \frac{1}{m-1/2}  , 
       \end{equation}
    which are just the raising and lowering operators (8) derived directly
    from the factorization of the Schr\"{o}dinger equation.

{\center \section*{III. 3D HYDROGEN ATOM}        }
~~
\par
    Now we address the Runge-Lenz vector for a 3D hydrogen atom, 
      \begin{equation}
        {\bf a=p \times l}-{\mit i}{\bf p-r}/{\mit r} .
      \end{equation}
   Defining $a_{\pm}=a_{x} \pm ia_{y}$,  we get
      \begin{equation}       
        a_{\pm}=\mp (\frac{\partial}{\partial x} \pm i\frac{\partial}
        {\partial y})(l_{z} \pm 1) \pm \frac{\partial}{\partial z}l_{\pm}-
        \frac{x \pm iy}{r} .
     \end{equation}
   In the spherical coordinate system $a_{\pm}$ may be expressed as
     \begin{eqnarray}    
       a_{\pm}&=&\pm \frac{\partial}{\partial r}[\cos\theta l_{\pm}-
       \sin\theta{\rm e}^{\pm i\phi}(l_{z}\pm 1)]  \nonumber \\
        & &\mp \frac{1}{r}[\sin\theta \frac{\partial}{\partial \theta}
         l_{\pm}\pm \cos\theta l_{\pm}(l_{z} \pm 1)\mp \sin\theta{\rm e}^
         {\pm i\phi}l_{z}(l_{z}\pm 1)]  \nonumber \\
        & &-\sin{\theta}e^{\pm i\phi} .
     \end{eqnarray}
 Similarly,
     \begin{eqnarray}
       a_{z}&=&p_{x}l_{y}-p_{y}l_{x}-ip_{z}-z/r \nonumber \\
        &=&\frac{\partial}{\partial r}[\frac{1}{2}\sin\theta
       ({\rm e}^{i\phi}l_{-}-{\rm e}^{-i\phi}l_{+})-\cos\theta] \nonumber \\        
        & &+\frac{1}{r}\{\sin\theta \frac{\partial}{\partial \theta}+
        \frac{1}{2}[\cos\theta(l_{+}l_{-}+l_{-}l_{+})-\sin\theta({\rm e}^
        {i\phi}l_{z}l_{-}+{\rm e}^{-i\phi}l_{z}l_{+})]\} \nonumber \\
        & &-\cos\theta .
     \end{eqnarray}
   (18) and (19) operating on the simultaneous eigenfunction of
   $(H, {\bf{l}}^{2}, l_{z}), \Psi_{nlm}(r, \theta, \phi)=R_{nl}(r)
   Y_{lm}(\theta, \phi)$, we get
     \begin{eqnarray}
       a_{\pm}\Psi_{nlm}&=&\pm \frac{d}{dr}R_{nl}(r)[(l+1)d_{l, \pm m}
       Y_{l+1, m\pm 1}+ld_{l-1, -(\pm m+1)}Y_{l-1, m\pm 1}]  \nonumber \\
                 & & \mp \frac{1}{r}R_{nl}(r)[l(l+1)d_{l, \pm m}Y_{l+1, m\pm 1}
      -l(l+1)d_{l-1, -(\pm m+1)}Y_{l-1, m\pm 1}] \nonumber \\
                 & &\pm R_{nl}(r)[d_{l, \pm m}Y_{l+1, m\pm 1}-d_{l-1, -(\pm m+1)}
       Y_{l-1, m\pm 1}] , \\
       a_{z}\Psi_{nlm}&=&-\frac{d}{dr}R_{nl}(r)[(l+1)c_{l, m}Y_{l+1, m}
      -lc_{l-1, m}Y_{l-1, m}]\nonumber \\
               & &+\frac{1}{r}R_{nl}(r)[l(l+1)c_{l, m}Y_{l+1, m}
       +l(l+1)c_{l-1, m}Y_{l-1, m}] \nonumber \\
               & &-R_{nl}(r)[c_{l, m}Y_{l+1, m}+c_{l-1, m}Y_{l-1, m}] , 
     \end{eqnarray}
   where
     \begin{equation}
       c_{l, m}=\sqrt{\frac{(l+1)^{2}-m^{2}}{(2l+1)(2l+3)}},  \quad
       d_{l, m}=\sqrt{\frac{(l+m+1)(l+m+2)}{(2l+1)(2l+3)}} .
     \end{equation}
   Using (20) and (21),  we may get
     \begin{eqnarray}
       l_{-}a_{+}\Psi_{nlm}&=&[\frac{d}{dr}-\frac{l}{r}+\frac{1}{l+1}]
        R_{nl}(r)(l+1)(l+m+2)c_{l, m}Y_{l+1, m} \nonumber \\
                 & &+[\frac{d}{dr}+\frac{l+1}{r}-\frac{1}{l}] 
       R_{nl}(r)l(l-m-1)c_{l-1, m}Y_{l-1, m}  , \\
       l_{+}a_{z}\Psi_{nlm}&=&-[\frac{d}{dr}-\frac{l}{r}+\frac{1}{l+1}]
       R_{nl}(r)(l+1)(l-m+1)d_{l, m}Y_{l+1, m+1} \nonumber \\
                 & &+[\frac{d}{dr}+\frac{l+1}{r}-\frac{1}{l}]
       R_{nl}(r)l(l+m)d_{l-1, -(m+1)}Y_{l-1, m+1}  , \\
       l_{-}a_{z}\Psi_{nlm}&=&-[\frac{d}{dr}-\frac{l}{r}+\frac{1}{l+1}]
       R_{nl}(r)(l+1)(l+m+1)d_{l, -m}Y_{l+1, m-1}\nonumber \\
                & &+[\frac{d}{dr}+\frac{l+1}{r}-\frac{1}{l}]
       R_{nl}(r)l(l-m)d_{l-1, m-1}Y_{l-1, m-1} .
     \end{eqnarray}
   Now, we may define the operators $S_{\pm}$ constructed from the Runge-Lenz
   vector and angular momentum operator, 
     \begin{eqnarray}
       S_{+}&=&l_{-}a_{+}+(l_{z}-l+1)a_{z} ,  \nonumber \\
       S_{-}&=&l_{-}a_{+}+(l_{z}+l+2)a_{z} .
     \end{eqnarray}
   It is easily verified that
     \begin{eqnarray}
       S_{+}\Psi_{nlm}&=&(\frac{d}{dr}-\frac{l}{r}+\frac{1}{l+1})
       R_{nl}(r)(l+1)(2l+1)c_{l, m}Y_{l+1, m}(\theta, \phi) \nonumber \\
               & &\propto a_{+}(l)R_{nl}(r)Y_{l+1, m}(\theta, \phi)  , \nonumber \\
       S_{-}\Psi_{nlm}&=&(\frac{d}{dr}+\frac{l+1}{r}-\frac{1}{l})
       R_{nl}(r)l(2l+1)c_{l-1, m}Y_{l-1, m}(\theta, \phi)   \nonumber \\
               & &\propto a_{-}(l)R_{nl}(r)Y_{l-1, m}(\theta, \phi)   , 
    \end{eqnarray}
   where
     \begin{equation}
      a_{+}(l)=(\frac{d}{dr}-\frac{l}{r}+\frac{1}{l+1}) , \quad
      a_{-}(l)=(\frac{d}{dr}+\frac{l+1}{r}-\frac{1}{l}) , 
     \end{equation}
   which are equivalent to the angular momentum raising and lowering operators
   $A_{\pm}(l)$ derived from the factorization of  radial Schr\"{o}dinger
   equation for a 3D hydrogen atom$^{[6]}$,   
     \begin{equation}
       A_{+}(l)=(\frac{d}{dr}-\frac{l+1}{r}+\frac{1}{l+1}) , 
       A_{-}(l)=(\frac{d}{dr}+\frac{l}{r}-\frac{1}{l}) , 
     \end{equation}
   which operate on the radial wavefunction $\chi_{nl}(r)=rR_{nl}(r)$. It is
   seen that the effect of $S_{+}(S_{-})$ is to increase (decrease) the angular
   momentum $l$ by $1$, but keep  the energy and magnetic quantum number
    $m$ unchanged. Therefore,  to clearly indicate the selection rules, 
    $S_{+}$ and $S_{-}$ may be relabelled as
     \begin{eqnarray}
      S_{+}\rightarrow S(n, l\uparrow, m)&=&l_{-}a_{+}+(l_{z}-l+1)a_{z}   , \nonumber \\
      S_{-}\rightarrow S(n, l\downarrow, m)&=&l_{-}a_{+}+(l_{z}+l+2)a_{z}  . 
     \end{eqnarray}
   Similarly,  using the Runge-Lenz vector and angular momentum operator, 
   one may construct the other two
   kinds of raising and lowering operators
     \begin{eqnarray}
      S(n, l\uparrow, m\uparrow)&=&(l_{z}+l-1)a_{+}-l_{+}a_{z} , \nonumber \\
      S(n, l\downarrow, m\uparrow)&=&(l_{z}-l-2)a_{+}-l_{+}a_{z} ,       \\
      S(n, l\uparrow, m\downarrow)&=&(l_{z}-l+1)a_{-}-l_{-}a_{z} ,  \nonumber \\
      S(n, l\downarrow, m\downarrow)&=&(l_{z}+l+2)a_{-}-l_{-}a_{z} .
    \end{eqnarray}

{\center\section*{IV. ISOTROPIC HARMONIC OSCILLATORS}  }
~~
\par
   It is  well-known that an {\sl n}D isotropic harmonic oscillator  has the
   dynamical symmetry $SU_{n}$. For a 3D isotropic harmonic oscillator, apart
   from the Hamiltonian $H$ and angular momentum {\bf l},  there exist five 
   additional conserved quantities which constitute a quadruple tensor
    \begin{eqnarray}
      Q_{xy}&=&xy+p_{x}p_{y}, Q_{yz}=yz+p_{y}p_{z}, 
      Q_{zx}=zx+p_{z}p_{x}, \nonumber \\
      Q_{1}&=&\frac{1}{2}[(x^{2}-y^{2})+(p^{2}_{x}-p^{2}_{y})]   ,  \\
      Q_{0}&=&\frac{1}{2\sqrt{3}}[(x^{2}+y^{2}-2z^{2})+(p^{2}_{x}+p^{2}_{y}
      -2p^{2}_{z})]  .  \nonumber
    \end{eqnarray}
   It can be shown that there exist four relations among the nine
   conserved quantities,  so a 3D isotropic harmonic oscillator is also a
   completely degenerate system and moves,  in general,  along an elliptic
   orbit and the direction of the semi-axes and eccentricity are
   characterized by the quadruple tensor. It has been shown$^{[6]}$ that
   the radial Schr\"{o}dinger equation of an isotropic harmonic oscillator
   as well as a hydrogen atom can be factorized. Nevertheless, it was noted that
   for an isotropic harmonic oscillator,  {\sl two} (rather than one) kinds of
   raising and lowering operators can be derived from  factorization, 
    and these operators themselves are {\sl not} conserved quantities (unlike
    the situation of the hydrogen atom).  However, it can be shown that in
    terms of the two kinds of raising and lowering operators one
   can construct the conserved quantities characterizing an elliptic orbit.

   For simplicity,  we take a 2D isotropic harmonic oscillator as example.
    A 2D isotropic harmonic oscillator has the dynamical symmetry $SU_{2}$, 
  which is locally isomorphic to $SO_{3}$ (the dynamical symmetry of a 2D hydrogen
  atom). For a 2D isotropic harmonic oscillator in classical mechanics, one
    may construct two conserved quantities, $Q_{xy}=xy+p_{x}p_{y}$ and
 $Q_{1}=\frac{1}{2}[(x^{2}-y^{2})+(p_{x}^{2}-p_{y}^{2})]$,  and the
 intersection angle $\gamma$ of the major axis with the $x$ axis is
 determined by $\tan2\gamma=Q_{xy}/Q_{1}$,  and the
 eccentricity$\propto[Q^{2}_{xy}+Q^{2}_{1}]^{\frac{1}{2}}$.
                                                     
    In ref. [6], it  was shown that from the factorization of radial
    Schr\"{o}dinger equation for a 2D isotropic oscillator, two kinds of
    raising and lowering operators,  $A_{\pm}$ and $B_{\pm}$,  operating on
    the radial wave function $\chi_{m}(\rho)$,  can be derived
     \begin{eqnarray}
        A_{\pm}(m)&=&\frac{d}{d\rho}\mp \frac{m \pm 1/2}{\rho}\pm \rho, 
          \nonumber \\
        B_{\pm}(m)&=&\frac{d}{d\rho}\mp \frac{m \pm 1/2}{\rho}\mp \rho, 
    \end{eqnarray}
   which may be replaced by $a_{\pm}(m)$ and $b_{\pm}(m)$ operating on $R_{m}(\rho)
  =\chi_{m}(\rho)/\sqrt{\rho}$, 
    \begin{eqnarray}
      a_{\pm}(m)&=&\frac{d}{d\rho}\mp \frac{m}{\rho}\pm \rho,    \nonumber \\
      b_{\pm}(m)&=&\frac{d}{d\rho}\mp \frac{m}{\rho}\mp \rho .
   \end{eqnarray}
  When operating on the whole wave function, ${\rm e}^{im\phi}
  R_{m}(\rho), a_{\pm}(m)$ and $b_{\pm}(m)$ may be replaced by $a_{\pm}$ and
   $b_{\pm}$, 
    \begin{eqnarray}
      a_{+}&=&e^{i\phi}[\frac{\partial}{\partial \rho}
        -\frac{1}{\rho}l_{z}+\rho] , \nonumber\\
      a_{-}&=&e^{-i\phi}[\frac{\partial}{\partial \rho}
        +\frac{1}{\rho}l_{z}-\rho]  , \nonumber \\
     b_{+}&=&e^{i\phi}[\frac{\partial}{\partial \rho}
     -\frac{1}{\rho}l_{z}-\rho] , \nonumber \\
     b_{-}&=&e^{-i\phi}[\frac{\partial}{\partial \rho}
     +\frac{1}{\rho}l_{z}+\rho]  .
   \end{eqnarray}
 It is easily shown that,  similar to (5) for a 2D hydrogen atom, 
  for a 2D isotropic oscillator we have
   \begin{equation}
     [l_{z}, a_{\pm}]=\pm a_{\pm}, [l_{z}, b_{\pm}]=\pm b_{\pm} , 
   \end{equation}
 i.e. both $a_{\pm}$ and $b_{\pm}$ are also angular momentum raising
 and lowering operators,  but unlike the hydrogen atom, 
  here $a_{\pm}$ and $b_{\pm}$ are not conserved quantities, 
    \begin{equation}
       [H, a_{\pm}]=\mp a_{\pm}, [H, b_{\pm}]=\pm b_{\pm} .
    \end{equation}
  Therefore,  the operators $a_{\pm}$ and $b_{\pm}$ themselves can not be
  directly equivalent to the conserved quantities characterizing the elliptic
  orbit. However,  it is easily verified that the conserved quantities
  $l_{z}, Q_{1}$ and $Q_{xy}$ can be constructed in terms of $a_{\pm}$ and
  $b_{\pm}$, 
       \begin{eqnarray}
       l_{z}&=&\frac{1}{4}(a_{-}a_{+}-b_{+}b_{-})   ,  \nonumber \\
       Q_{1}&=&-\frac{1}{4}(a_{-}b_{-}+b_{+}a_{+})  ,  \nonumber \\
       Q_{xy}&=&-\frac{i}{4}(a_{-}b_{-}-b_{+}a_{+}) .
    \end{eqnarray}

{\center \section*{V. SOME DISCUSSION ABOUT 1D SYSTEMS}          }
~~
\par
  The Schr\"{o}dinger's factorization method and the concept of raising
  and lowering operators were extended extensively in supersymmetic quantum
  mechanics$^{[13-16]}$ to treat the Schr\"{o}dinger equation for a particle
   in a general 1D potential  $V(x)$. It was shown that for a potential
   $V(x)$,  provided the ground bound  state energy $E_{0}$ is finite
   ($E_{0}\ne -\infty$) and the ground state wave  function $\Psi_{0}(x)$
   is differentiable,  one can construct the corresponding raising and
   lowering operators,  $A^{+}$ and $A$,  and the Schr\"{o}dinger equation
   can always be factorized. The supersymmetric partner Hamiltonian, 
    $H_{-}=A^{+}A$ and $H_{+}=AA^{+}$,  have the same energy spectra, 
     $E^{(+)}_{n}=E^{-}_{n+1}, (n=0, 1, 2, \cdots, )$,  except the ground state
     energy of $H_{-}$  $(E^{(-)}_{0}=0)$,   and the eigenstates with the
      same eigenvalue of $H_{-}$ and $H_{+}$ are connected with each
      other by $A^+$ and $A$. It was shown that this is due to the shape
      invariance$^{[17]}$ of $V(x)$,  which may also be considered
      as a special kind of dynamical symmetry.

   It is interesting to note that the classical orbits of all bound
    particles in a regular 1D potential are always closed. However,  no 
    classical close orbits exist for some
    singular potentials,  e.g. ,  for
    a 1D hydrogen atom ($e=1$)$^{[17]}$
      \begin{eqnarray*}
       V(x)= -\frac{1}{|x|}, (-\infty <x <+\infty) .
     \end{eqnarray*}
   On the other hand,  for a quantum 1D hydrogen atom,  the ground state
   energy $E_{0}=-\infty$,  and $\Psi_{0}(x)\sim \sqrt{\delta(x)}$ 
   is not differentiable at the origin. Thus,  it seems understandable 
   why the Schr\"{o}dinger equation of a 1D hydrogen atom can not be
   factorized.

   As for a 1D harmonic oscillator,  the energy eigenvalues and eigenstates
   are well known,  which are quite similar to those for a 3D isotropic 
   harmonic oscillator ($l=0$ case). The raising and lowering operators
   derived from factorization are $a^{+}=\frac{1}{\sqrt{2}}(x-\frac{d}{dx})$
    and $a=\frac{1}{\sqrt{2}}(x+\frac{d}{dx})$,  which connect the neighboring 
    eigenstates with opposite parity ($\Delta N=1$). However,  
    it should be noted that the 1D harmonic oscillator potential 
    formally corresponding to a 3D isotropic oscillator ($V(r)=r^{2}/2,  r\geq 0$)
    is
     \[ V(x)= \left \{
       \begin{array}{ll}   
       x^{2}/2 & x\geq 0 \\
       \infty & x< 0   
     \end{array} 
     \right. \quad , \]
   whose energy levels are $E_{N}=(N+1/2)$,   $N=1, 3, 5, \cdot$. The usually   
    adopted 1D harmonic oscillator is $V(x)=x^{2}/2$ ($-\infty <x<+\infty$) 
    with reflection symmetry,  whose levels are $E_{n}=N+1/2$,   $N=0, 1, 2, 3, 
    \cdot$ and the neighboring eigenstates are of opposite parity. From this,  
    one can understand why there exist two kinds of raising and lowering operators, 
    $A_{\pm}$ and $B_{\pm}$,  for an $n$D ($n\geq 2$) isotropic harmonic oscillator, 
    and $A_{\pm}$ and $B_{\pm}$ are different in form from the operators 
    $a^{+}$ and $a$ for a 1D harmonic oscillator. However,  one may use the 
    product operator of $A$ and $B^{[6]}$, i.e. the operator $C$ as the raising
    and lowering operators connecting the neighboring eigenstates with the 
    {\sl same} parity. In fact,  for a 3D isotropic harmonic oscillator$^{[6]}$
      \begin{eqnarray*}
        C(l=0, N\uparrow \uparrow)=\frac{d^{2}}{dr^{2}}+r^{2}
          -2r\frac{d}{dr}-1
      \end{eqnarray*}
   is the same form as $2a^{+}a=\frac{d^{2}}{dx^{2}}+x^{2}-2x\frac{d}{dx}-1$   
   for a 1D harmonic oscillator,  and both have the selection rule $\Delta N=2$.

   This work is supported by the National Natural Science Foundation of China and 
   the Doctoral Program Foundation of Institute of Higher Education of China.

\end{document}